\documentclass[epj]{svjour}
\newcommand*{\epem}{\ensuremath{\mathrm{e^{+}e^{-}}}}
\newcommand*{\ep}{\ensuremath{\mathrm{e^{\pm}p}}}
\newcommand*{\pp}{\ensuremath{\mathrm{pp}}}

\newcommand*{\tautau}{\ensuremath{\tau^{+}\tau^{-}}}

\newcommand*{\sqrs}{\ensuremath{\sqrt{s}}}
\newcommand*{\lumu}{\ensuremath{\mathrm{10^{34}/cm^{2}s^{-1}}}}
\newcommand*{\lumi}{\ensuremath{\mathrm{ab^{-1}}}}
\newcommand*{\IntL}{\ensuremath{\int\cal{L}}}
\newcommand*{\InstL}{\ensuremath{\cal{L}}}
\newcommand*{\allevts}{\ensuremath{\mathrm{all\ events}}}

\newcommand*{\Zrun}{\ensuremath{\mathbf{Z}}}
\newcommand*{\WWrun}{\ensuremath{\mathbf{WW}}}
\newcommand*{\ZHrun}{\ensuremath{\mathbf{ZH}}}
\newcommand*{\ttrun}{\ensuremath{\mathbf{t\bar{t}}}}
\newcommand*{\edm}{\ensuremath{\mathtt{EDM4hep}}}
\newcommand*{\kfh}{\ensuremath{\mathtt{key4hep}}}
\newcommand*{\fccsw}{\ensuremath{\mathtt{FCCSW}}}
\newcommand*{\raw}{\ensuremath{\mathtt{RAW}}}
\newcommand*{\aod}{\ensuremath{\mathtt{AOD}}}
\newcommand*{\geant}{\ensuremath{\mathtt{Geant4}}}
\newcommand*{\delphes}{\ensuremath{\mathtt{DELPHES}}}
\newcommand*{\hepspec}{\ensuremath{\mathtt{HEPSpec}}}
\newcommand*{\compe}{\ensuremath{\mathtt{CE}}}
\newcommand*{\kkmcee}{\ensuremath{\mathtt{KKMCee}}}
\newcommand*{\pythia}{\ensuremath{\mathtt{Pythia8}}}
\usepackage{graphics}
\usepackage[T1]{fontenc}
\usepackage{textgreek}

\begin{document}
\title{Offline Computing resources for FCC-ee and related challenges}
\author{Cl\'ement Helsens\inst{1} \and Gerardo Ganis\inst{1}% etc
}                     % Do not remove
\offprints{}          % Insert a name or remove this line
\institute{CERN}
\date{{\it (Submitted to EPJ+ special issue: A future Higgs and Electroweak factory (FCC): Challenges towards discovery, Focus on FCC-ee)}}
\abstract{
The international Future Circular Collider (FCC) study aims at designing \pp, \epem, \ep\ colliders to be built in a new 100~km tunnel in the Geneva region. The electroweak, Higgs and top factory (FCC-ee) is designed to provide collisions at a centre of mass energy range between 90 (Z-pole) and 365~GeV ($\mathrm{t\bar{t}}$) and unprecedented  integrated luminosities, producing huge amounts of data which will pose significant challenges to data processing.
In this essay we discuss the needs in terms of storage and CPU for the diverse phases of the project, and the possible solutions mostly based on the models developed for HL-LHC.
%Heterogeneous computing/Data centres/Cloud?)
    %
\PACS{
      {PACS-key}{describing text of that key}   \and
      {PACS-key}{describing text of that key}
     } % end of PACS codes
} %end of abstract
\maketitle
\section{Introduction}
\label{intro}
The FCC-ee, first stage of the Future Circular Collider (FCC) integrated programme~\cite{FCCPhys}, plans to collide \epem at various centre of mass energies. The nominal run plan for expected instantaneous and integrated luminosities and relevant events statistics for the different physics runs is reported in Table~\ref{table:runplan}.

\begin{table}[h!]
\caption{Run plan for FCC-ee. Source~\cite{FCCPhys}.}
\label{table:runplan}
\centering
\begin{tabular}{ | c | c | c | c | c |}
\hline
 Run  & \sqrs\ (GeV) & \InstL\  (\lumu) & \IntL\  (\lumi) & Statistics \\ 
\hline
 \Zrun  & 91.2    & 230      & 150     & $3\cdot10^{12}\ \mathrm{Z}$ decays (visible)\\ 
 \WWrun & 160     & 28       & 12      & $10^{8}\ \mathrm{W^+W^-}$ \ events \\  
 \ZHrun & 240     & 8.5      & 5       & $10^{6}\ \mathrm{ZH}$ events \\
 \ttrun & 350,~365 & 1.8,~1.55 & 0.2,~1.5 & $10^{6}\  \mathrm{t\bar{t}}$\ events \\
\hline
\end{tabular}
\end{table}

FCC-ee is planned to start operation after the high luminosity stage of LHC (HL-LHC) is completed, i.e. around 2040. Offline computing at FCC-ee will therefore take advantage of the HL-LHC computing model and achievements. The computing needs for FCC-ee are driven by the \Zrun\ run and are usually considered comfortable, in particular considering that no or negligible pile-up is expected for an \epem\ collider\footnote[1]{Machine--Detector--Interface induced backgrounds can potentially be important at FCC-ee; they are the subject of ongoing detailed studies and the current results show that they should not significantly affect the size of the data samples~\cite{Voutsinas2020}.}.
% We need also to define a reference time scale and reference use-cases.
The exercise we are discussing in the essay relates to the preparation of the Feasibility Study Report, FSR, to be submitted to the next European Strategy Update. We assume the bulk of the studies, driven by the Physics Performance group~\cite{fcceepp}, will be run during the three years 2022-2024. We also need assumptions for the number of detector concepts to be evaluated. This is more complicated, and the only possible approach is to estimate the resources needed as a function of the number of detector variations to be evaluated. 

% In the next sections we will investigate more in depth the meaning of all this in terms of storage and computing requirements for the diverse phases of the project, namely Monte Carlo generation, detector simulation, event reconstruction and data analysis. A concrete use-case to confront with is the preparation of the Feasibility Study report~(FSR).

The essay is structured as follows.
After presenting the typical workflows which we consider relevant for this study in Section~\ref{sec:workflows}, in Sections~\ref{sec:storage}~and~\ref{sec:computing} we estimate the needs in terms of storage and computing, for the diverse phases of the project, namely Monte Carlo generation, detector simulation, event reconstruction and data analysis. In Section~\ref{sec:summary}, we discuss the estimates and outline the main areas which we consider challenging. In Section~\ref{sec:conclusions} we sum up and conclude.

\section{Typical workflows}
\label{sec:workflows}

The resource requirements obviously depend on the objectives to be pursued, which in turn, determine the workflow to be followed.
We can initially distinguish the following general cases:
\begin{enumerate}
    \item Collision data reconstruction and analysis. This concerns running experiments or test-beam data processing. The reconstruction part is typically well defined and run only a few times (for re-calibration purposes or improved reconstruction algorithms). The analysis part is by nature chaotic and less standard, although it may contain well defined phases, for example the preparation of tuples for the final selections and fits
    \footnote[2]{The preparations of the analysis tuples by individuals or working groups have acquired an increasingly important role in HEP experiments not only in an attempt to homogenise the set of high-level variables to work with, but also to optimise the use of resources. In particular optimisation of data I/O, a known bottleneck, may possibly be achieved by exploiting the experiment's own infrastructure. An example of this is the case of the ALICE analysis trains~\cite{ALICEtrains}, which minimise I/O by applying a set of registered algorithms to the same event readout only once.}.
    \item Full/fast Monte Carlo simulations, including digitisation, followed by reconstruction and analysis. This concerns current experiments and experiments being designed. Interpretation of test-beam data may also require simulation, at least to some extent. For running experiments the reconstruction and analysis phase are the same as for collision data.
    The amount of simulated data required depends on the use-case. For running or test-beam experiments, the amount of simulated data should be enough to make the associated statistical uncertainty component negligible. At LEP, a rule of thumb of ten times more Monte Carlo data than collision data was often used. In general this would not be  applicable at LHC, given the amount of resources taken by the simulation, except perhaps for studying specific background features in reduced portions of the phase space.
    \item Parameterised Monte Carlo simulations, possibly followed by the analysis. This typically concerns experiments in the design stage, although use of these techniques to interpolate between full simulation parameter points is not uncommon for collision data analysis, especially in searches for new physics. As in the previous case, the amount of parameterised simulated data should be enough to make the associated statistical uncertainty component negligible. 
\end{enumerate}
For each of these cases we have to consider a number of variations resulting from the physics studies - or several detector concepts, for experiments being designed - with different resource requirements. Good organisation of the different activities can certainly optimise the requirements, in particular by eliminating duplications. 

\section{Storage}
\label{sec:storage}
The storage needs depend on the data format, and the persistency and redundancy requirements for the data. For a project at the design stage, redundancy is mostly connected to the optimisation of other resources, network and CPU (it might be, for example, more efficient to duplicate some data locally than to access them remotely); it will depend on the resources finally made available and their geographical distribution.
Data persistency is also connected with the available resources and with the trade off between the cost of recreating the data and the cost of storing them. For example, there is the tendency to keep the data used for a publication as a reference, although what is strictly needed is the recipe to reproduce them. Efficient bookkeeping of the configuration settings and software environments used for creating a data set would certainly allow the needs to be better balanced. 

The data format depends on the {\it event data model}; different phases of the experiment will require different levels of detail and, in the initial steps, possibly different data structures. However, as soon as we come to describing the {\it physics content} of an HEP event, a set of standard observables can be defined.

In the computing model being setup for FCC, based on the \fccsw~\cite{FCCSW,FCCSW2}\ framework, data at any level are described by the data structures provided by \edm~\cite{EDM4hep}, a common event data model developed for future HEP  experiments. This means that
\begin{itemize}
    \item Full/fast simulation generates an \edm\ output;
    \item Reconstruction algorithms understand \edm\ input and write \edm\ output;
    \item Parameterised simulation produces an \edm\ output where the quantities have the same meaning as those from reconstruction. High level reconstruction algorithms, such as vertex finding, should be applicable both to reconstruction output and parameterised simulation output;
    \item Analysis is run on \edm\ files; in particular the same analysis algorithms should be applicable to fully simulated and reconstructed events and to parameterised events resulting from parameterised simulation. 
\end{itemize}

In \fccsw, full and fast simulation refer to the full and fast mode of \geant~\cite{GEANT}, and includes also digitisation. The reconstruction algorithms are FCC-specific or taken from \kfh~\cite{key4hep,key4hep2}. Parameterised simulation is obtained with \delphes~\cite{DELPHES}. 

\subsection{\raw\ event sizes}
\label{sec:raweventsizes}
The \raw\ format is the event format used to describe collision data and fully simulated data. The exact format is only available once the detector design choices are frozen. To estimate the \raw\ event sizes for experiments in the design phase, a baseline solution for a typical detector is needed. For FCC-ee there are currently two such baseline solutions under study: CLD, an adaptation of the CLIC baseline detector; and IDEA, a new innovative detector concept for \epem\ colliders. These two detector concepts are being studied in some details. Table~\ref{table:rawcldidea} summarises the understanding at the time of writing based on~\cite{Grancagnolo}.   

\begin{table}[h!]
\centering
\caption{Typical \raw\ event sizes in kB for the \Zrun\ run for the two baseline detector solutions~\cite{Grancagnolo} and the ALEPH detector~\cite{ALEPHInNumbers}; the contribution of the final states originating from the Z exchange (Z decays) is singled out from the expected total (\allevts). }
\label{table:rawcldidea}
\begin{tabular}{ | c | c | c | c | c | c | c |}
\hline
                & \multicolumn{2}{c|}{CLD}
                & \multicolumn{2}{c|}{IDEA} & \multicolumn{2}{c|}{ALEPH} \\
\hline
Readout channels& \multicolumn{2}{c|}{1.9 G}
                & \multicolumn{2}{c|}{2.8 G} & \multicolumn{2}{c|}{1 M} \\
\hline
Sub-detector    & Z decays & \allevts
                & Z decays & \allevts & Z decays & \allevts \\
\hline
Vertex          & 1.3~kB & 62~kB & 1.3~kB & 62~kB & & \\
Tracker         & 1.4~kB & 102~kB & 500~kB & 595~kB & &\\
Calorimeter     & 230~kB & 920~kB~(*) & 500~kB & 2000~kB~(*) & &\\
Muon            & 0.03~kB & 0.75~kB & 0.03~kB & 0.75~kB & &\\
\hline
Total           & 233~kB & 1085~kB & 1001~kB & 2658~kB & 120~kB & 550~kB\\
\hline
\multicolumn{7}{l}{(*) For the calorimeters, reference~\cite{Grancagnolo} does specify numbers for the \allevts\ case, only for}\\
 \multicolumn{7}{l}{the IDEA pre-shower; the numbers are obtained by applying the same factor 4 expected for the} \\ 
\multicolumn{7}{l}{IDEA pre-shower to all the calorimeters.}
\end{tabular}
\end{table}
It is evident from Table~\ref{table:rawcldidea} that the technology choice can make a difference and more refined/innovative technologies may result in a very large amount of data. Ongoing software developments indicate that this is potentially problematic not only for the storage, but also for the computing needs of simulation and/or reconstruction. The numbers for the IDEA tracker, a high granularity stereo drift chamber, already include optimisation based on the use of FPGA to reduce the data sample by a factor 15~\cite{Grancagnolo}. While there is a general belief that there is still room for improvement, if only by applying standard techniques such zero suppression, a range of 1$-$2~MB seems appropriate for the study at hand.

\subsection{\aod\ event sizes}
\label{sec:aodeventsizes}
Analysis Data Objects format, or \aod, is the format used for analysis, therefore the output created by the reconstruction phase or by parameterised simulation. Table~\ref{table:eventsizes} shows the typical event sizes for different types of events processed through \delphes\ in \edm\ format. This is expected to be a good estimation of the typical event sizes after reconstruction of collision data or fully simulated events. From this table, a range of 5$-$10~kB per event seems appropriate for this study.

\begin{table}[h!]
\caption{Typical FCC-ee event sizes, in kB, for different types of events processed through \delphes{} in \edm{} format (using the IDEA detector concept card with track covariance~\cite{DelphesIDEA}). The values measured in ALEPH are taken from Ref.~\cite{ALEPHInNumbers} and shown for comparison.$^3$}
%\footnotemark[3] }
\label{table:eventsizes}
\centering
\begin{tabular}{ | c | c | c |}
\hline
 Process  & \sqrs\ (GeV) & Average size\ (kB)   \\ 
\hline
 $\mathrm{Z\to u\bar{u},d\bar{d},s\bar{s}}$  & 91.2 &  4.9 \\ 
 $\mathrm{Z\to c\bar{c}}$  & 91.2    & 5.2 \\ 
 $\mathrm{Z\to b\bar{b}}$  & 91.2    & 5.5 \\
 $\mathrm{Z\to \tautau}$   & 91.2    & 1.2 \\
 \hline
 Z decays, ALEPH, AOD  & 91.2 & 9.0 \\
 Z decays, ALEPH, MINI  & 91.2 & 2.2 \\
 \hline
 ZH inclusive & 240 & 8.9 \\
 ZZ inclusive & 240 & 6.6 \\
 $\mathrm{W^+W^-}$\ inclusive & 240 & 6.4  \\
 \hline
 $\mathrm{t \bar t}$ & 350 & 13 \\
 \hline
\end{tabular}
\end{table}

\footnotetext[3]{The ALEPH MINI format only contained the information need for end-user analysis~\cite{ALEPHOffline}, in compressed form.}

\subsection{Storage requirements}
\label{sec:storageneeds}
The rough estimates for the event sizes provided in the previous sections can be used to estimate the amount of storage required at various stages.

\subsubsection{\raw\ data and the event format for full simulation}
\label{sec:rawstorageneeds}
The amount of \raw\ data expected for the FCC-ee runs based on the estimates discussed in Section~\ref{sec:raweventsizes} are given in Table~\ref{table:rawdata}.

\begin{table}[h!]
\caption{\raw\ data estimates for FCC-ee.}
\label{table:rawdata}
\centering
\begin{tabular}{ | c | c | c | c |}
\hline
 Run  & \sqrs\ (GeV) & Statistics & \raw\ data\\ 
\hline
 \Zrun  & 91.2   &  $3\cdot10^{12}$ $\mathrm{Z}$ decays (visible) & 3--6 EB \\ 
 \WWrun & 160     &  $10^{8}$ $\mathrm{W^+W^-}$ events & 0.1--0.2 PB \\  
 \ZHrun & 240     &  $10^{6}$ $\mathrm{ZH}$ events & 1--2 TB\\
 \ttrun & 350, 365 &  $10^{6}$ $\mathrm{t \bar t}$ events & 1--2 TB\\
\hline
\end{tabular}
\end{table}

As expected, the \Zrun\ run is by far the most demanding in terms of storage and will be used as a reference in the following. It can be seen from the table, that the \Zrun\ run values are in the range of a few EB, i.e. of the order of the values expected for HL-LHC~\cite{Boccali}. By the time FCC-ee is brought to operation, not before 2040, the amount of  \raw\ experiment collision data should not therefore present a challenge, and should be manageable with a simple evolution of the HL-LHC model.

% The picture is somehow different when the full simulation needs are considered.
% The \raw\ data event sizes are also an estimation of the average size of fully simulated events. In order to evaluate the full potential of a detector choice, samples of simulated data with a statistical power similar to the expected data might be required, and this for several diversification of the detector solutions. This of course represents a potential challenge for the FSR preparation phase which is the main target of this essay. A good and efficient strategy to optimize the storage needs with an interplay between fast and full simulation is definitely required.

The picture is somehow different when the full simulation needs are considered.
To understand if a detector choice has the potential to match the FCC-ee requirements in terms of systematic uncertainty control, very large samples of simulated data might be required, and this for several diverse detector solutions. However, as mentioned at the beginning of Section~\ref{sec:storage}, the persistency requirements of the simulated data are different to those of the collision data, availability of the simulated data being strictly needed only for the time required by reconstruction runs.
So, if the storage of \raw\ simulated data is potentially a challenge for the FSR preparation phase, an efficient strategy to optimise the storage needs over time will provide a means to mitigate the impact on resource requirements. This strategy should allow for the interplay between fast and full simulation.

% Whatever the case, the storage of \raw\ simulated data can represents a potential challenge for the FSR preparation phase, and a good and efficient strategy to optimize the storage needs, including interplay between fast and full simulation, is definitely required.

\subsubsection{\aod\ data samples}
\label{sec:aodstorageneeds}
Based on the estimates discussed in Section~\ref{sec:aodeventsizes}, the amount of \aod\ data expected for the FCC-ee runs is given in Table~\ref{table:aoddata}.
The amount of data expected for the \Zrun\ run is of the order of tens of PB, which represents a considerable amount during the FSR preparation phase, requiring a dedicated strategy and resource management.

\begin{table}[h!]
\caption{\aod\ data estimates for FCC-ee.}
\label{table:aoddata}
\centering
\begin{tabular}{ | c | c | c | c |}
\hline
 Run  & \sqrs\ (GeV) & Statistics & \aod\ data\\ 
\hline
 \Zrun  & 91.2   &  $3\cdot10^{12}$ $\mathrm{Z}$ decays (visible) & 15$-$30 PB \\
 \WWrun & 160     &  $10^{8}$ $\mathrm{W^+W^-}$ events & 0.5$-$1 TB \\  
 \ZHrun & 240     &  $10^{6}$ $\mathrm{ZH}$ events & 5$-$10 GB\\
 \ttrun & 350,~365 &  $10^{6}$ $\mathrm{t \bar t}$ events & 5$-$10 GB\\
\hline
\end{tabular}
\end{table}

\section{Computing resources}
\label{sec:computing}
Estimating computing resources is more complicated because more unknowns, such as the evolution of the efficiency of the various codes, enter the game.

The metrics for computing resources that is generally accepted is \hepspec. Exact numbers of \hepspec\ provided by a Computing Element (\compe) depends on the detailed hardware configuration, which is impossible to know at this stage. In the following we will assume that one core of today's CERN OpenStack \compe\ brings 10-15 \hepspec.
The current computing resources assigned to FCC at CERN amount to 9000 \hepspec\, which we will also refer to as a {\it computing unit}.

The $\mathrm{q \bar q}$\ events at the \Zrun\ run centre-of-mass energies are typically the most demanding case and will be used as a reference, together with the only currently published example of an FCC-ee case study full analysis involving rare b-mesons decays~\cite{Amhis:2021cfy}. This can offer indications of the impact of the reconstruction and/or the analysis phases on the computing resources required.

% The benchmark codes we have chosen are
% \begin{enumerate}
%     \item Full simulation of the FCC-hh baseline detector; this serves as upper conservative limit.
%     \item Full simulation of the simplified IDEA tracker and vertex detector in FCC-ee.
%     \item Parametrized simulation of $Z$ hadronic decays through \delphes.
%     \item Final phase (fitting, plotting) of a typical analysis in \edm\ format. 
% \end{enumerate}
% We are using \qq\ events at the \Zrun\ as reference.

The numbers shown in the remainder of the essay have been obtained by running benchmark codes on a CERN Openstack machine with 16 cores, 32 GB RAM~\cite{Openstack}.

% \subsection{Reference time scale and use-cases}
% \label{sec:refs}
% We need also to define a reference time scale and reference use-cases. The exercise we are discussing in the essay relates the preparation of the Feasibility Study Report, FSR, to be submitted to the next European Strategy Update. We assume the bulk of the studies, driven by the Physics Performance group~\cite{fcceepp}, will be run during the three years 2022-2024. We also need assumptions for the number of detector concepts to be evaluated. This is more complicated, and the only possible approach is to estimate the resources needed as a function of the number of detector variations to be evaluated. 
\subsection{Monte Carlo generation}
\label{sec:compmcgen}
The real time taken for the generation of 100k $\mathrm{q \bar q}$ events with \pythia~\cite{Pythia8} and with \kkmcee~\cite{KKMCee} at the \Zrun\ run centre-of-mass energies is shown in Table~\ref{table:gen}; the two reference generators give similar results. For comparison, the table also shows numbers for the generation of \tautau\ and \textmu$^+$\textmu$^-$ events with \kkmcee, which, per event, are up to a factor two larger than those for $\mathrm{q \bar q}$.

\begin{table}[h!]
\caption{Time estimated to generate $\mathrm{q \bar q}$, \tautau\ and \textmu$^+$\textmu$^-$ events at the Z peak.}
\label{table:gen}
\centering
\begin{tabular}{ | c | c | c | c | c |}
\hline
 Generator  & Process & 100k/core & \Zrun\ sample/core & \Zrun\ sample/9000 HS06\\ 
\hline
 \pythia  & $\mathrm{q \bar q}$ & 148~s  &  $4\cdot10^{9}$~s = $\sim$126~y & 50--75 days \\
 \kkmcee & $\mathrm{q \bar q}$ & 151~s   &  $4\cdot10^{9}$~s = $\sim$126~y & 50--75 days \\  
\kkmcee & \tautau & 195~s   &  $0.25\cdot10^{9}$~s = $\sim$8~y & 3--4.5 days \\  
\kkmcee & \textmu$^+$\textmu$^-$ & 334~s   &  $0.44\cdot10^{9}$~s = $\sim$14~y & 5--7.7 days \\  
\hline
\end{tabular}
\end{table}
Table~\ref{table:gen} also shows the time required to generate a sample equivalent to that expected from the \Zrun\ run  on a single CERN core and using the computing resources currently assigned to FCC at CERN. We have already seen that this step is challenging and full scale production requires an optimised use of resources. Of course an efficiency optimisation of the Monte Carlo codes is also a possibility to be taken into account.

One additional comment relates to the use of a dedicated generator for the decay of heavy quarks, such as EvtGen~\cite{evtgen}, which was used for the analysis in Ref.~\cite{Amhis:2021cfy}. These analyses require exclusive samples, currently obtained by filtering away unwanted decays. Since the number of events to be generated is not large, the inefficiency of this technique is not
currently a limitation, but it represents a waste of CPU usage for rare hadrons (such as B$_c$) as most of the events generated are skipped. Improvements in the efficiency of the filtering technique are open areas of work. 

% In these cases there is a need to produce exclusive samples through a filtering technique which, for some rare processes, may be quite inefficient. 
% While the a

% the generation needs to be repeated a lot of times before the desired decay is achieved. While currently this inefficiency does not represent a limitation, as the number of events to generate is not that large, but it represent a waste of CPU usage as most the generated events are skipped. 

\subsection{Detector simulation}
\label{sec:compsim}

The full simulation time per event in the CLD detector is approximately 20 seconds per hadronically decaying \Zrun\ boson; the same number for IDEA cannot be derived yet. The CLD number is similar to the time required to simulate $\mathrm{t\bar{t}}$ events at ATLAS or CMS once the average multiplicity scaling is taken into account, so it can be considered a realistic estimation of what can be done with current simulation techniques\footnote[4]{Any improvement in the speed of simulation coming from the current extensive fast simulation R\&D is extremely welcome, although unlikely to change these numbers significantly in the timescale of the FSR.}.
% For comparison, the simulation of a $q\bar{q}$ event with the ALEPH Geant3 based simulation program on today CERN typical hardware takes on average 0.3 s. ({\bf Can we explain the scaling?}).  
Table~\ref{table:sim} shows the projected integral time estimates.
Considering the full statistics at the \Zrun\ it becomes clear that the current computing resources are largely insufficient as it would take 2 to 3 thousand years to simulate it.

As can be seen in the table, the computing resources to simulate a full statistics equivalent of the \Zrun\ data sample in 2-3 years (the expected duration of \Zrun\ and also the preparation time for the FSR) is about 10 million \hepspec, which is of the order of the resources available to the LHC experiments today~\cite{ATLASneeds}.

\begin{table}[h!]
\caption{Time estimated to simulate $\mathrm{q \bar q}$ events at the Z peak.}
\label{table:sim}
\centering
\begin{tabular}{ | c | c | c | c |}
\hline
 Process & 1k/core & \Zrun\ sample/core & \Zrun\ sample/9000 HS06\\ 
\hline
  $\mathrm{q \bar q}$ & 20k sec = 5h33min  &  $6\cdot10^{13}$~s = $\sim1.9\cdot10^{6}$~y & 2.1--3.2$\cdot10^{3}$~y \\
\hline
\end{tabular}
\end{table}

This can be seen in two ways. On one side, producing the full statistic samples during the FCC-ee operations, although resource demanding, is not likely to be an issue; the computing resources available to FCC-ee will be at least equivalent to those available for HL-LHC. On the other side, for the FSR, it will certainly be impossible to have full statistics samples in full simulation of all the detector concepts.

When the fast simulation option is enabled in Geant4, the response of the sub-detectors is parameterised and the particle transport simplified. CMS has shown that applying these techniques to the calorimeter system, an acceptable precision in the description of the detector response can be kept with an overall speed-up by a factor of about 10. This is still not enough for full statistics samples for the FSR, but it goes in the right direction and the FCC community is certainly looking with interest at the Geant4 team efforts to improve the quality and speed of the fast simulation option. 
% An alternative to simulate all the physical processes could be using LHCb methodology of partial detector simulation, \emph{e.g.} not to simulate Cherenkov processes for calorimeters when the physics channel studies does not use that information. 

% Finally, a potentially interesting area of development for FCC is the use of deep learning techniques for detector simulation, which starts to show promising results at the LHC~\cite{atlfast3}. 

Alternative approaches to reduce the impact of detector simulation on the overall simulated event processing budget, include methodologies of partial detector simulation, such the one adopted by LHCb, e.g. not to simulate Cherenkov processes when the physics channel studies do not use that information, or new approaches to detector simulation, such as those based on deep machine learning technologies, which start to have a role at the LHC~\cite{atlfast3}, with promising results.

% Considering a scenario where we simulate the events while we take data at the \Zrun\, this gives us 3 years to simulate all what we need. Consequently, the computing resources we would need are about 9 millions HS06 per year, which represents the current annual total needs of an experiment like ATLAS~\cite{ATLASneeds}.

\subsection{Reconstruction}
\label{sec:compreco}

The event reconstruction is expected to be less busy at FCC-ee than at FCC-hh and (HL-)LHC, particularly for the tracks as the multiplicity is orders of magnitude smaller than at the LHC, thus greatly simplifying the pattern recognition. For comparison, at ALEPH the reconstruction step took about 10-15\% of the simulation time~\cite{ALEPHInNumbers}, while at Belle-II, it accounts for about a third of the total processing time~\cite{Belle2} and is dominated by tracking and depends on the amount of background. For FCC-ee it is therefore reasonable to assume that the reconstruction time could potentially lead to a maximum comparable to half the simulation time discussed in Section~\ref{sec:compsim}.

\subsection{Detector parameterisation}
\label{sec:compparam}
FCC studies use \delphes\ for a fast parameterised simulation of the detector concepts.  
The \delphes\ processing of 100k $\mathrm{q \bar q}$ events generated with \pythia\ at the \Zrun\ takes 212 seconds on a single core CERN machine. Table~\ref{table:param} shows that using the CERN computing unit, between 2.5 and  4 months would be needed to produce the full \Zrun\ statistics.

\begin{table}[h!]
\caption{Time estimated to simulate $\mathrm{q \bar q}$ events at the Z peak with \delphes{}}
\label{table:param}
\centering
\begin{tabular}{ | c | c | c | c |}
\hline
 Process & 100k/core & \Zrun\ sample/core & \Zrun\ sample/9000 HS06\\ 
\hline
  $\mathrm{q \bar q}$ & 212~s  &  $6.4\cdot10^{9}$~s = $\sim202$~y & 0.22--0.34~y \\
\hline
\end{tabular}
\end{table}

\subsection{Analysis}
\label{sec:companalysis}
Quantifying the needs for physics analyses depends on the use case. To illustrate it we will focus on one recently published analysis using all the common tools~\cite{Amhis:2021cfy}. This example focuses on precisely measuring a rare b-hadron decay, and  tight cuts needs to be applied to achieve excellent signal purity. In order to achieve an accurate estimation of the backgrounds, it was not possible to generate the expected inclusive data statistics. About 10 billion events, including exclusive decays with larger acceptance, were generated and reconstructed with \delphes{} in the \edm{} data format occupying approximately 50 TB of disk space. Analysing these events to create small ntuples with all the heavy calculations done (vertexing, candidate building etc...) with the current CERN FCC batch resources takes half a day. The second step with the final selection can be achieved locally within less than an hour.
% How the scaling to full data statistic 

\section{Ways ahead}
\label{sec:summary}

Table~\ref{table:summary} summarises the number of events which could be produced per day with one computing unit and with the equivalent of the current ATLAS computing resources.

\begin{table}[h!]
\caption{Number of $\mathrm{q \bar q}$ events that can be produced per day with one computing unit and with the equivalent of the ATLAS computing resources.}
\label{table:summary}
\centering
\begin{tabular}{ | c | c | c | c | c |}
\hline
     & Generation & Simulation & Reconstruction & \delphes\ \\ 
\hline
 Computing unit & 3.5--5.2$\cdot10^{10}$ & 2.6--3.9$\cdot10^{6}$ &  5.2--7.8$\cdot10^{6}$ & 2.4--3.6$\cdot10^{10}$ \\
\hline
\hline
 ATLAS equivalent & 3.5--5.2$\cdot10^{13}$ & 2.6--3.9$\cdot10^{9}$ &  5.2--7.8$\cdot10^{9}$ & 2.4--3.6$\cdot10^{13}$ \\
\hline
\end{tabular}
\end{table}

In projecting the numbers summarised in Table~\ref{table:summary} we have to consider the two cases of the FCC-ee operation and of the preparation of the FSR separately, as already done in some cases above. Since the FCC-ee operation will come after the HL-LHC experience, there is little doubt that the resources required, both in terms of storage and of computing, should be affordable.

The situation is different concerning the studies for the FSR. At the time of writing, the Physics Performance group has 33 case studies to be addressed. Projecting the resources needed by one of these cases~\cite{Amhis:2021cfy}, and assuming that all that can be shared between case studies is effectively shared, i.e. that all removable duplications are removed, it is probably safe to say that a close-to-full expected statistic is possible at the parameterised simulation level. However, a full simulation for each detector concept would not be possible.

Based on these considerations we see that the following main areas should be addressed: improvement of the parameterised simulation, the interplay of  full/fast/parameterised simulations; the minimal needs in terms of simulation statistics.

\subsection{Improving the parameterised simulation}
\label{sec:summaryparm}

The tracking description of the parameterised simulation with \delphes{} has been considerably improved during the studies following the publication of the FCC CDR. A detailed fast simulation of the tracks, including track covariance, allows much more detailed and realistic studies of tracking algorithms and therefore of observables related to tracking, such as vertexing or heavy flavour tagging. This has required the introduction of geometrical concepts in \delphes{}, though possibly in a simplified version. The question is if the same kind of approach could be used for other parts of the detector. The obvious first candidate is the calorimeter, where full simulation results, including spacial development of showers, could be parameterised and applied directly at \delphes{} level. Similar approaches could be envisaged for other detectors, such as muon chambers, Cherenkov detectors or inner detectors. The separation barrel/forward and insensitive region (cracks) simulation could potentially be improved without a large impact on the processing time. Parameterised simulation of multiple scattering (for charged particle propagation to the calorimeters), at least in the inner detectors could also be implemented, as well as parameterisation for detached vertexes. 

\subsection{Interplay of full/fast/parameterised simulations}
\label{sec:summaryinter}
Somehow connected with the improvements of the parameterised simulation discussed in Section~\ref{sec:summaryparm} is the interplay between the different levels of simulation. If resources are only available for limited studies in full simulation it is important at best to use these studies to feed the better understanding back to parameterised simulation, or to derive the results of the studies. These techniques were already used for the CDR to understand the needs in terms of detector performance for a given measurement: the relevant detector response was studied in full simulation and translated into the impact on the result with parameterised simulation~\cite{CDRinterplay}.

\subsection{Minimal needs in terms of simulation statistics}
\label{sec:summaryneeds}
In Section~\ref{sec:workflows} it was mentioned that a rule of thumb for the requirements of the case studies in terms of the simulation statistics was {\it at least equivalent to the expected data sample}. We saw in the previous sections how difficult is to have samples satisfying this requirement for detailed simulations. There is therefore the need to go beyond the rule of thumb and develop systematic evaluation technologies which could be statistically more powerful, thereby reducing the number of events required.
Developing these could also be very useful for the analysis of the collision data, when they come. 

% \subsection{Approaches to sustainable computing}
% Even if it seems there are no limitations with current computing ressources available at LHC, we should nevertheless significantly contribute to continuously improve code efficiency to improve the execution time, CPU usage by , disk usage by storing only the relevant objects. 

\section{Conclusions and outlook}
\label{sec:conclusions}

In this essay we have started investigating the computing needs of FCC-ee in view of the next phases of the project and operation. We have shown how, probably without surprise, these requirements are dominated by the \Zrun\ run. We have also shown that despite being large, the requirements for storage and computing resources should not pose problems during the operation of the machine, planned to start after HL-LHC. Given this timescale there is the possibility to benefit from all the advances, developments and findings of HL-LHC, including the resource sustainability aspects.
Finally we have shown how the preparation of the FSR for the next European Strategy Upgrade is potentially challenging, and will require the experiment groups to develop ways to optimally use and manage the data samples available, {\it de facto} increasing their statistical power and reducing the effective resource needs. After all, scarcity of resources is often behind the birth of brilliant ideas.

\end{document}